# Strain Induced Slater Transition in Polar Metal LiOsO$_3$

Y. Zhang,[1] J. J. Gong,[1] C. F. Li,[1] L. Lin,[1] Z. B. Yan,[1] Shuai Dong,[2,*] and J.-M. Liu[1,3]

[1]*Laboratory of Solid State Microstructures, Nanjing University, Nanjing 210093, China*

[2]*School of Physics, Southeast University, Nanjing 211189, China*

[3]*Institute for Advanced Materials, Hubei Normal University, Huangshi 435002, China*

[*]Email: sdong@seu.edu.cn

[**Abstract**] LiOsO$_3$ is the first experimentally confirmed polar metal. Previous works suggested that the ground state of LiOsO$_3$ is just close to the critical point of metal-insulator transition. In this work the electronic state of LiOsO$_3$ is tuned by epitaxial biaxial strain, which undergoes the Slater-type metal-insulator transition under tensile strain, i.e., the G-type antiferromagnetism emerges. The underlying mechanism of bandwidth tuning can be extended to its sister compound NaOsO$_3$, which shows an opposite transition from a antiferromagnetic insulator to a nonmagnetic metal under hydrostatic pressure. Our work suggests a feasible route for the manipulation of magnetism and conductivity of polar metal LiOsO$_3$.

## I. INTRODUCTION

It is well known that metals cannot exhibit ferroelectric distortions because of the screening effects of the conduction electrons. However, in 1965, Anderson and Blount suggested that the loss of inversion symmetry could occur in metallic materials through a continuous structural transition so that polar metals could exist [1]. The first success in the experimental finding of polar metal is LiOsO$_3$, which was reported to undergo a second-order structural phase transition from the centrosymmetric structure (space group $R\bar{3}c$) to the non-centrosymmetric structure (space group $R3c$) at temperature $T$=140 K with metallic behavior unchanged [2].

The novel properties of polar metal have attracted much attention since the first experimental report about LiOsO$_3$ in 2013. The major concerns of LiOsO$_3$ are not only the potential applications but more importantly possible couplings between the ferroelectricity and metallicity which are usually mutually exclusive. Many efforts have been devoted into understanding the origin of the ferroelectric-like distortion in the metallic state of LiOsO$_3$, and the microscopic mechanism has been investigated [3-6]. Although several previous experimental and theoretical studies have demonstrated the primary issues of polar metal, the physics of $LiOsO_3$ is still puzzling and controversial [2,7-9].

Previous works based on both local density approximation (LDA) + $U$ and LDA + dynamic meaning field theory (DMFT) both suggested that the ground state of $LiOsO_3$ is just located near the critical point of metal-insulator transition [10]. If the Coulomb repulsion of $Os^{5+}$ is weak enough, the strong hybridization between Os's $5d$ and O's $2p$ orbitals and spin-orbit coupling (SOC) will suppress the local magnetic moment in $LiOsO_3$ completely and then the ground state should be metallic. But if the Coulomb repulsion is significant, a G-type antiferromagnetic (G-AFM) structure would be established with every nearest-neighboring spins oppositely aligned, and meanwhile $LiOsO_3$ will become an insulator. For the half-filling $t_{2g}$ orbital of $Os^{5+}$, the Hubbard $U$ will ideally split the $t_{2g}$ orbitals into full-filled lower Hubbard bands and empty upper Hubbard bands, with the opposite signs of periodic potential on each nearest neighbor, i.e. the Slater-type metal-insulator transition [see Fig. S1 in Supplementary Materials (SM)] [11].

Up to now all experimental evidences showed that $LiOsO_3$ should be a nonmagnetic (NM) weak-correlated metal, which is different from other $5d^3$ osmium oxides such as $NaOsO_3$, $Cd_2OsO_7$, etc [12-14]. As clarified in Ref. [10], the main factor that makes $LiOsO_3$ different from $NaOsO_3$ is the difference between their lattice structures. The Os-O-Os network is more compact in $LiOsO_3$ than that in $NaOsO_3$, so the $5d$-$2p$ hybridization is stronger and the effective hopping between the nearest-neighbor Os's $5d$ orbitals is larger in $LiOsO_3$ than that in $NaOsO_3$. Hence, the band widths of $5d$ orbitals in $LiOsO_3$ are relatively wider than those in $NaOsO_3$. Then the narrower $5d$ bands of $NaOsO_3$ favor a stable G-AFM structure.

Then it is natural to expect that the suppressed magnetic ordering of $LiOsO_3$ could be revived by reducing the kinetic energy of $5d$ electrons which enhances the electron localization. Puggioni *et al.* proposed a strategy of electronic structure control in $LiOsO_3$ by enhancing the electronic correlations in the $LiOsO_3$ layers of an ultrashort period $LiOsO_3$/$LiNbO_3$ superlattice, and their calculation showed that the insulating and magnetic state of $LiOsO_3$ could be driven by the reduction of bandwidth of $t_{2g}$ orbitals in the superlattice geometry [15].

In this manuscript, we will consider an simpler route, i.e. to apply biaxial strains, to tune the distances between ions in the Os-O-Os network of $LiOsO_3$. Then under the tensile case, the itinerant electrons should be more localized and the magnetic order is expectable. Although the strain effects were addressed before, the magnetism was neglected [16,17]. Using the first-principles density functional theory (DFT), our calculation reveals that $LiOsO_3$ will turn into G-AFM insulator with the unchanged crystal symmetry under tensile biaxial strain. Our results suggest an alternative route for the manipulation of the physical properties and functionality of LiOsO$_3$.

## II. COMPUTATION METHODS

The DFT calculations are performed using the pseudo-potential plane wave method as implemented in Vienna *ab initio* Simulation Package (VASP) [18-20]. The electron interactions are described using the Perdew-Burke-Ernzerhof (PBE) of the generalized gradient approximation (GGA) [21]. The projected augmented wave (PAW) [22] pseudo-potentials with a 500 eV plane-wave cutoff are used, including three valence electrons for Li ($1s^22s^1$), nine for Na ($2s^22p^63s^1$), fourteen for Os ($5p^66s^25d^6$), and six for O ($2s^22p^4$). Starting from the experimental structures, the lattice constants and all atomic coordinates are fully relaxed within the initial space group, until the Hellman-Feynman forces on every atom are converged to less than 1.0 meV/Å. A 13×13×7 mesh for the cell of LiOsO$_3$ in $R3c$ hexagonal phase (containing 30 atoms), and a 13×13×13 mesh for the unit cell of NaOsO$_3$ are used for the Brillouin-zone sampling.

The Coulomb repulsion in correlated electron systems is usually characterized by the on-site Hubbard $U$. Since the Coulomb repulsion should be insignificant in 5$d$ systems because of the spatially extended 5$d$ orbitals, we have carefully investigated the ground state of LiOsO$_3$ by performing two different types of DFT + $U$ methods [10]: the LDA + $U$ method introduced by Liechtenstein *et al.* in which the exchange splitting $J$ and the Hubbard $U$ are considered separately and the simplified LSDA (local spin density approximate) + $U$ method introduced by Dudarev *et al.* which only needs a parameter $U_{eff} = U-J$ [23,24]. In the LDA + $U$ case, the suitable value of Hubbard $U$ in LiOsO$_3$ should be in the range of 1.0~2.0 eV; but in LSDA+$U$ case, the value of $U_{eff}$ should be small enough (~0 eV), i.e. a bare LSDA calculation is a proper choice.

In our calculations, the space groups are kept unchanging upon the external strain or pressure, i.e. structural transition is not taken into account although it may occur in some cases [25,26]. To guarantee this assumption, the dynamic stability of lattice will be checked using its phonon spectrum as well as the elastic coefficients [27] (Fig. S2 in SM). The ferroelectric polarization for the insulating state is calculated using the Berry phase method [28,29]. The total spontaneous polarization $P$ for a given crystalline symmetry can be calculated as the sum of ionic and electronic contributions.

## III. RESULTS AND DISCUSSION

We take the hexagonal cell of LiOsO$_3$ in the non-centrosymmetric phase of $R3c$ as the reference structure, as shown in Fig. 1(a). First, previous works have proved that the magnetic phase of LiOsO$_3$ could only be NM or G-AFM whatever the condition is [3,9,10,15]. So in the following we only consider the competition between the NM phase and G-AFM phase. The total and partial density of states (DOS) are shown in Fig. 1(b-c). As expected, LiOsO$_3$ is insulating in the G-AFM state while the NM state is metallic. For both cases, Os's 5$d$ orbitals and O's 2$p$ orbitals are overlapped around the Fermi level, suggesting the strong hybridization between them. Due to the crystalline electric field, the Os's 5$d$ orbitals are split into the $t_{2g}$ triplet manifold and $e_g$ doublet manifold. Thus the magnetic configuration will be formed by the total moment of the half-filled $t_{2g}$ orbitals of Os$^{5+}$.

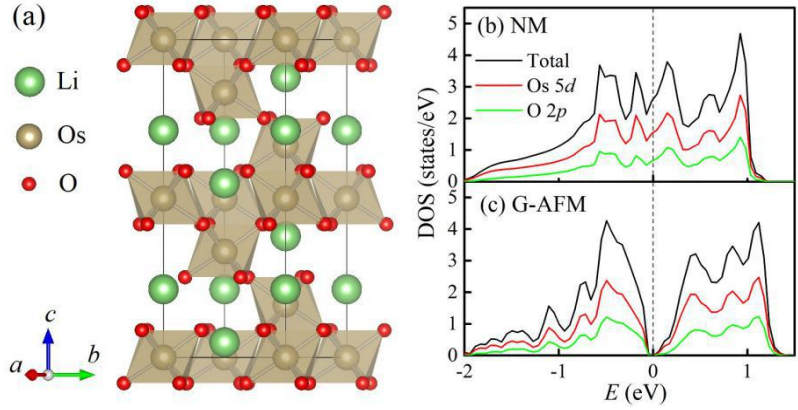

**Figure 1.** (a) The hexagonal crystal structure of LiOsO$_3$ in non-centrosymmetric phase. The total and partial density of states (DOS) of LiOsO$_3$ (per f.u.) in (b) NM state and (c) G-AFM state, calculated by LSDA.

In order to reduce the $t_{2g}$ bandwidth of Os$^{5+}$ via making Os-O-Os network looser, we consider to apply biaxial strain which can be achieved in experiments by epitaxial growth techniques. In crystals with $R3c$ space group, the lattice parameter $a$ equals to the lattice parameter $b$. The $a/b$ plane biaxial strain is defined as $\varepsilon=(a-a_0)/a_0$, where $a_0$ is the optimized lattice parameter for the fully relaxed LiOsO$_3$. We study the ground state of LiOsO$_3$ under an appreciable range of strains, from -5% to 5%, while the negative numbers represent compressive strain and the positive numbers represent tensile strain. With constrained $a/b$, the out-of-plane lattice parameter $c$ and internal atomic coordinates are fully relaxed, and the crystal symmetry of the LiOsO$_3$ cell is kept unchanged at the $R3c$ space group during the optimized process (whose stability is further verified, see Fig. S2 in SM).

To accurately confirm the magnetic order of LiOsO$_3$ under biaxial strains, first we perform the LSDA + SOC calculation on LiOsO$_3$, here the spin-orbit coupling (SOC) is considered because that several works have found that SOC may play a significant role in the magnetic properties of 5$d$ osmium oxides [14,30-32]. We compare the total energy of LiOsO$_3$ cell in the NM state and G-AFM state. The results are summarized in Fig. 2(a-c). We can see that the energy difference between the total energy of LiOsO$_3$ cell in the NM state and G-AFM state varies with the magnitude of strain. As shown in Fig. 2(a), the total energy of LiOsO$_3$ in the NM state would gradually go beyond that in the G-AFM state as the magnitude of strain increasing, which means the ground state of LiOsO$_3$ turn from the NM phase to G-AFM phase. Within a certain range of tensile strain (~>2%), the G-AFM state would be the most stable for LiOsO$_3$. In other cases, the ground state of LiOsO$_3$ would remain in the NM state. In practice, the possible substrate candidates may be LiTaO$_3$ (~2.15% larger) or the (111) surface of LaAlO$_3$ or YAlO$_3$ (6.1% or 3.8 % larger). The magnetic moment and band gap are also growing with the tensile strain, as presented in Fig. 2(b-c), indicating that the G-AFM phase becomes more and more stable as the magnitude of tensile strain increases.

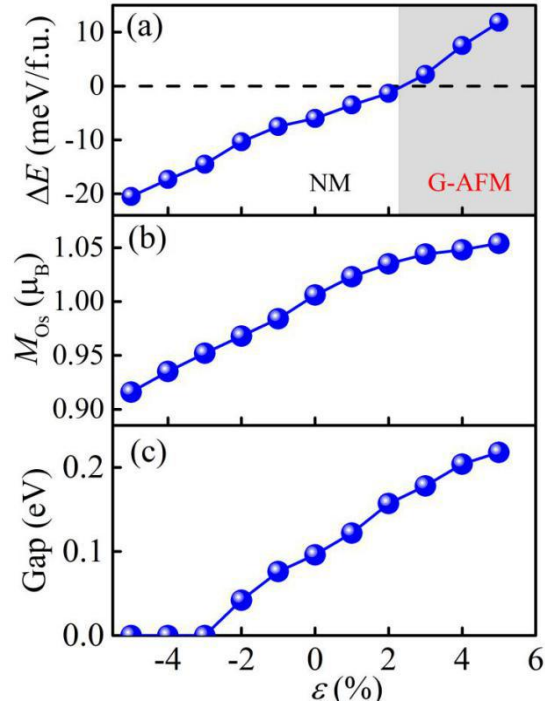

**Figure 2.** Results of LSDA+SOC calculations for LiOsO$_3$ as a function of biaxial strain $\varepsilon$. Negative values represent compressive strains, and positive values represent tensile strains. (a) The energy difference between the G-AFM state and NM state ($\Delta E = E_{NM} - E_{G\text{-}AFM}$). (b) The magnetic moment of Os ion in the G-AFM state. (c) The band gap of LiOsO$_3$ in the G-AFM state.

In order to reveal the underlying physical mechanism, the calculated structural parameters of LiOsO$_3$ as a function of strain are shown in Fig. S3 in SM. Under the tensile condition, the nearest-neighbor Os-Os distance (as well as Os-O-Os bond angle) becomes larger due to the expanded in-plane lattice constant although the out-of-plane lattice constant is shrunk. The compressive strain leads to the opposite tendency. Obviously, the longer Os-O bond and the larger distance between the nearest-neighbor Os-Os pair reduce the orbital hybridization and thus the bandwidth.

To verify the mechanism of tensile strains, the partial DOS of 5$d$ orbitals of each Os in the G-AFM state is plotted in Fig. 3(a). Here two cases are compared: unstrained and +3% tensile strained. As expected, the band width of $t_{2g}$ orbitals in the latter case is narrower that in the former case. So when the magnitude of tensile strain go beyond the critical point, the 5$d$ electrons of Os$^{5+}$ would tend to be localized and prefer to establish the G-AFM structure, which agrees with our expectation.

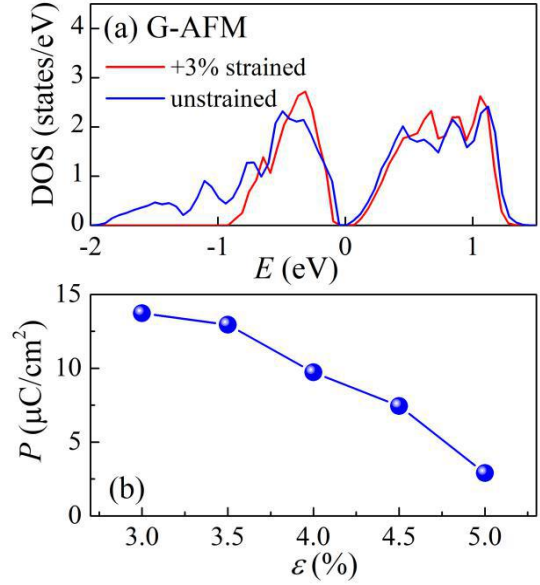

**Figure 3.** (a) The partial DOS of each Os in the G-AFM state of LiOsO$_3$ (per f.u.): unstrained and +3% tensile strained, calculated by LSDA. (b) The calculated ferroelectric polarization ($P$) along the $c$-axis as a function of biaxial strain $\varepsilon$.

The linear optical properties LiOsO$_3$ in the G-AFM state under tensile strains is also studied as a fringerprint of electronic structure. The calculated linear optical absorption coefficients before phase transition (undistorted) and after phase transition (+3% tensile strain) are shown in Fig. S4 in SM. Two main absorption peaks exist: one is located between 1.0 eV and 1.3 eV, the other is located between 3.5 eV and 4.5 eV. The first absorption peak is probably caused by electron transition from O's 2$p$ orbital below Fermi level to Os's $t_{2g}$ orbital, and the second absorption peak may be caused by electron transition from O's 2$p$ orbital to Os's $e_g$ orbital. The valleys located between 2.2 eV and 2.5 eV should be attributed to the gap between $e_g$ orbital and $t_{2g}$ orbital. It is obvious that the location of absorption peak is red-shifting under tensile strain, indicating that the energy level of Os's 5$d$ orbital is decreasing. This feature confirms that tensile strain could make the bandwidth of 5$d$ orbitals narrower as we expected.

In addition to above LSDA calculations, we also perform LDA+$U$ calculation on strained LiOsO$_3$ (the magnitude of strain is set to +3%) to confirm that our results is not dependent on particular computing method. We set $J/U = 0.25$ and restrict the range of $U$ in 1.0~2.0 eV. The energy difference between the G-AFM phase and NM phase and the magnetic moment of Os$^{5+}$ are shown in Fig. S5 in SM as a function of $U$. In the LDA+$U$ calculation, the tensile strain can enlarge the energy difference between G-AFM phase and NM phase and increase the local magnetic moment. Hence, both the LSDA calculation and LDA+$U$ calculation reach the same conclusion.

Since the G-AFM structure could open a gap in the Fermi level of LiOsO$_3$, we can calculate the spontaneous ferroelectric polarization in the non-centrosymmetric LiOsO$_3$ cell. The ferroelectric polarization is calculated using the Berry phase approach. The polarization is presented in Fig. 3(c) as a function of strain, which decreases with the increasing magnitude of strain. Clearly, the polar displacements in LiOsO$_3$ will be reduced by tensile biaxial strains. This result is easy to be understood since the polar displacements in LiOsO$_3$ are predominantly along the *c*-axis [2], and thus the reduced lattice constant along the *c*-axis suppresses the polarization.

Consequently, since we can manipulate the ground state of LiOsO$_3$ by applying tensile strain, then could we change the G-AFM insulating phase in NaOsO$_3$ into NM metallic phase via compressive strain considering the similarities of their structures? Our calculation indeed predicts that such magnetic transition can occur when the biaxial compression is beyond -8% (Fig. S6 in SM), which may be too large in practice.

Considering the feasibility for experiment, we consider the hydrostatic pressure on NaOsO$_3$ crystal instead, which can reach a similar effect. We optimize the structure of NaOsO$_3$ under pressures and then study the magnetic and electronic properties at the ground state. As shown in Fig. 4(a), the ground state of NaOsO$_3$ would turn into NM metallic phase under a relatively large pressure (>35 GPa) while the crystal symmetry is kept unchanged (whose dynamic stability is further verified according to the elastic coefficients. See SM for more details). Even for the G-AFM state, the pressure can suppress the local magnetic moment, as shown in Fig. 4(b). We compare the partial DOS of undistorted cell with the one under 50 GPa, as shown in Fig. 4(c), which indicates that the band widths of *5d* orbitals are clearly wider under pressure. That means the applied hydrostatic pressure makes the Os-O-Os network more compact so that the *5d* electrons will be more delocalized and the effective hopping between nearest-neighbor Os-Os pair's *5d* orbitals are larger, eventually lead to the quenching of the local moment of Os$^{5+}$ (*5d$^3$*) in NaOsO$_3$.

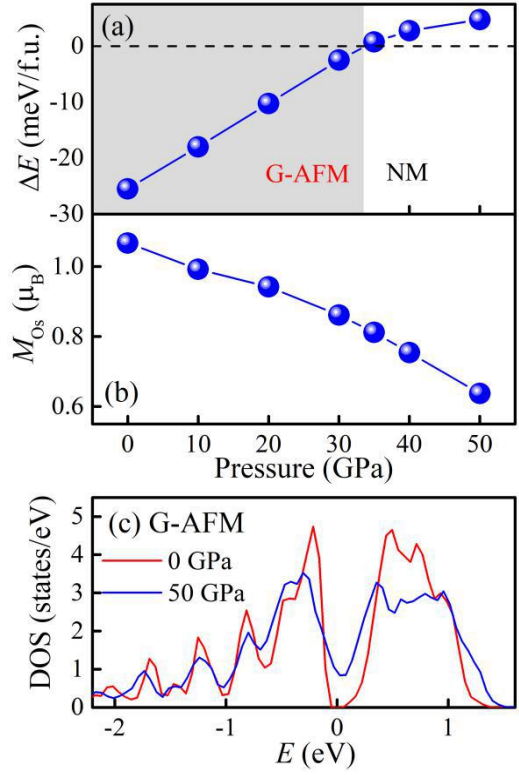

**Figure 4.** Results of LSDA+SOC calculations for NaOsO$_3$ as a function of applied hydrostatic pressure and partial DOS patterns in two cases. (a) The energy difference between NM state and G-AFM state ($\Delta E = E_{\text{G-AFM}} - E_{\text{NM}}$). (b) The magnetic moment of Os ion in the G-AFM state. (c) Partial DOS (per f.u.) of Os's $5d$ orbitals for the original one and the compressed one with 50 GPa hydrostatic pressure.

## IV. CONCLUSION

In summary, we have performed first-principles calculations to study the effects of biaxial strains on LiOsO$_3$. It is revealed that the ground state of LiOsO$_3$ would change from NM metal into G-AFM insulator under tensile biaxial strain. This magnetic phase transition would cause a Slater-type metal-insulator transition. The physical mechanism for the strain-induced phase transition in LiOsO$_3$ can also be applied to NaOsO$_3$. Oppositely, the ground state of NaOsO$_3$ would turn from G-AFM insulator into NM metal under sufficiently large hydrostatic pressure. Our results suggest a feasible way to manipulate the electronic and magnetic properties of polar metal LiOsO$_3$. Our work would motivate new applications of polar metal materials.

**ACKNOWLEDGMENTS**

This work was supported by the National Natural Science Foundation of China (11834002, 11674055, 51431006, 51721001) and the National Key Projects for Basic Researches of China (2016YFA0300101 and 2015CB654602).